\documentclass{article}
\usepackage[T1]{fontenc}
\usepackage[utf8]{inputenc}
\usepackage{ismir} 
\usepackage{amsmath,cite,url}
\usepackage{graphicx}
\usepackage{color}
\usepackage{booktabs}
\usepackage{multirow}
\usepackage{caption}
\title{TUTTI: Toward generalizable audio-to-score transcription via fully synthesized data}

\multauthor
  {Jianhuai Hu$^{*\dagger 1}$ \quad Yashan Wang$^{*1}$ \quad Shangda Wu$^{*3}$}
  {{\bf Zhancheng Guo$^{1}$ \quad Shijie Liang$^{1}$ \quad Wuna Meng$^{4}$ \quad Chuanqi Yang$^{5}$} \\
  {\bf Xiaobing Li$^{1}$ \quad Feng Yu$^{1}$ \quad Maosong Sun$^{\dagger 1,2}$}\\
  $^1$ Central Conservatory of Music \quad $^2$ Tsinghua University \quad $^3$ Independent Researcher \\
  $^4$ Wuhan Conservatory of Music \quad $^5$ Shanghai Conservatory of Music \\
  {\tt\small $^*$Equal contribution \quad $^\dagger$Corresponding author}\\
  {\tt\small hujianhuai@mail.ccom.edu.cn, sms@tsinghua.edu.cn}
  }

\def\authorname{J. Hu, Y. Wang, S. Wu, Z. Guo, S. Liang, W. Meng, C. Yang, X. Li, F. Yu, and M. Sun}

\usepackage{xspace} 
\newcommand{\tutti}{TUTTI\xspace}
\newcommand{\tutticorpus}{TuttiCorpus\xspace}
\begin{document}

\maketitle

\begin{abstract}
Generalizable Audio-to-Score (A2S) transcription is fundamentally constrained by the severe scarcity of high-quality, real-world paired data. Relying solely on existing human-annotated datasets often restricts the generalization of A2S models, limiting their efficacy primarily to single-instrumentation domains. To break this dependency on scarce real-world data, we introduce \textbf{\tutti} (\textbf{T}ransformer for \textbf{U}nified audio-\textbf{T}o-score \textbf{T}ranscription trained on Synthetic multi-\textbf{I}nstrumentation Data), a pre-training paradigm driven by a purely synthetic, large-scale dataset. Rather than using human-composed scores, we leverage a symbolic music generation model to generate a massive, highly scalable multi-instrumentation corpus and create audio-score pairs with expressive acoustic characteristics. Capitalizing on the generated data, we employ a standard Transformer encoder-decoder architecture. We empirically demonstrate that pre-training a unified attention-based model on generated, multi-instrumentation data yields a consistently stronger foundational representation than single-instrumentation training. When fine-tuned with downstream real-world datasets, \tutti outperforms previous approaches, establishing new overall state-of-the-art results across various A2S baselines. Notably, \tutti shows remarkable cross-instrument transferability, effectively adapting to unseen instruments with highly competitive performance. The source code and the TuttiCorpus dataset will be made publicly available at \url{https://github.com/a-musiclover/TUTTI}.
\end{abstract}

\section{Introduction}\label{sec:introduction}

Audio-to-Score (A2S) transcription aims to extract score-level symbolic representations (e.g., \texttt{ABC notation}, \texttt{**Kern}) directly from acoustic audio signals. Currently, A2S serves a dual purpose: it remains an indispensable tool for real-world music pedagogy and performance, while simultaneously acting as a pivotal data curation pipeline for generative music models~\cite{zeng2024end,alfaro2024transformer,martinezsevilla23_interspeech, soulxsinger}. As the demand for large-scale audio-score paired datasets grows, the ability to automatically transcribe performances into formats such as ABC notation becomes a key enabler for training generative models. Although early approaches decomposed A2S into isolated subtasks such as harmony estimation and note quantization, the paradigm has increasingly shifted towards end-to-end learning~\cite{cogliati2016transcribing,nakamura2018towards,shibata2021non,carvalho2017towards,roman2018end,roman2019holistic,liu2021joint}. However, despite recent architectural advances such as hierarchical decoding~\cite{zeng2024end} and customized Transformers~\cite{alfaro2024transformer}, the development of a unified foundation model remains constrained by several fundamental challenges.

\textbf{Data Scarcity Bottleneck} Data availability is the most critical hurdle in current A2S research~\cite{morsi2022bottlenecks,zeng2024end}. The total corpus of human-composed music is historically finite. Compounding this limitation, high-quality real-world acoustic performances aligned perfectly with their symbolic scores are exceedingly rare and expensive to annotate~\cite{asap-dataset,martinezsevilla23_interspeech}. Consequently, relying on existing human-annotated datasets makes it nearly impossible to train an A2S model capable of tackling the vast complexity of diverse, real-world musical compositions. 

\textbf{Lack of Generalizability across Instrumentations} Driven directly by this data scarcity, existing A2S models are highly fragmented and task-specific. The field has largely focused on the ``solo'' scenario—models strictly tailored to single-instrumentation transcriptions such as solo piano, string quartets, saxophone, etc.~\cite{zeng2024end,alfaro2024transformer,martinezsevilla23_interspeech,chen2019data}. To the best of our knowledge, a unified, generalizable A2S model capable of robustly transcribing musical pieces across a wide and unconstrained variety of instrumentations has not yet been proposed.

\textbf{Task-Specific Architectural Complexity} Given the historical lack of diverse, large-scale training data, the A2S field has naturally gravitated towards task-specific hybrid architectures. A common paradigm involves the use of convolutional neural networks (CNNs) as acoustic feature extractors, coupled with recurrent neural networks (RNNs) or attention mechanisms for sequence decoding~\cite{liu2021joint,alfaro2024transformer,zeng2024end,arroyo2022neural}. These hybrid pipelines are highly effective in data-scarce regimes because the strong local inductive biases of CNNs help prevent overfitting on small datasets~\cite{d2021convit, shao2022transformers}. However, this reliance on specialized acoustic front-ends inherently increases architectural complexity. This raises an underexplored question in the A2S domain: With the introduction of massive, large-scale synthetic training data, is it possible to bypass customized acoustic front-ends and rely entirely on a unified, standard architecture?

To answer this, we propose a paradigm shift from model-centric engineering to a data-centric approach. In this paper, we introduce \tutti, a pure Transformer-based encoder-decoder framework designed to perform generalizable multi-instrumentation A2S transcription. Treating A2S fundamentally as an ``audio-to-text'' sequence-to-sequence translation task~\cite{sutskever2014sequence}, \tutti directly decodes audio spectrogram features into ABC notation. Inspired by the success of pure attention-based models in Natural Language Processing~\cite{vaswani2017attention, radford2019language, raffel2020exploring,liu2024deepseek,guo2025deepseek} and their recent adaptations in symbolic music~\cite{wu2024melodyt5,wang2025notagen,wu2023tunesformer}, we intentionally adopt a standard Transformer architecture. By incorporating techniques such as interleaved ABC notation and hierarchical patch-level and char-level decoders~\cite{wu2024melodyt5,wu2023clamp,wang2025notagen}, pure Transformer architectures empowered by massive synthetic data can achieve highly competitive representations without strictly requiring customized CNN front-ends, highlighting that data scale is the primary driver of performance. Naturally, traditional CNN-based approaches stand to benefit similarly from such large-scale pre-training.
To overcome the data bottleneck, we pre-train \tutti on a massive synthetic dataset. Rather than relying on limited human compositions, we leverage NotaGen~\cite{wang2025notagen} to procedurally create over 363,000 unique, multi-instrumentation scores in ABC notation, which are subsequently rendered into audio. By pre-training on this unconstrained corpus, TUTTI learns a powerful foundational representation of multi-instrument acoustics.

The key contributions of our paper are as follows:
\begin{itemize}
    \item \textbf{\tutti}, an attention-based generalizable encoder-decoder model that directly transcribes audio spectrograms into ABC notation without relying on traditional CNN feature extractors.
    \item \textit{\tutticorpus}, a dataset of over 363,000 unique multi-instrumentation audio-score pairs. It enables the model to learn complex polyphonic structures across diverse musical textures through a purely generative data pipeline.
    \item Our experimental results demonstrate that \textbf{\tutti} achieves state-of-the-art performance across various instrumentations. Furthermore, we show that multi-instrumentation pre-training consistently outperforms single-domain training, and that the model exhibits generalizability to the unseen instrument (i.e., saxophone) which is entirely absent from the pre-training data.
\end{itemize}

\begin{figure*}
  \centering
  \includegraphics[alt={Architectural overview of the TUTTI model},width=0.9\linewidth]{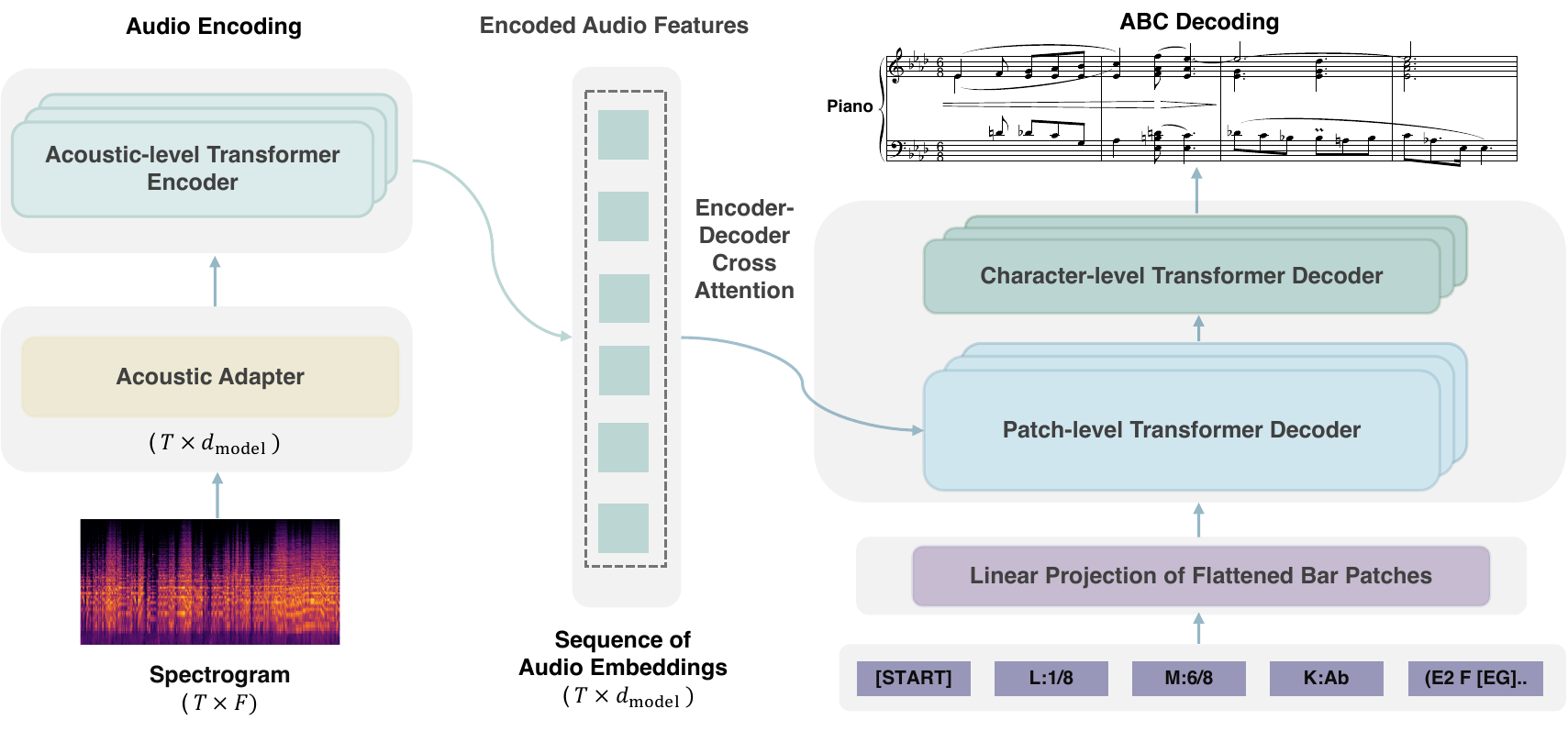}
  \caption{Architectural overview of \tutti. The model bridges the gap between continuous audio spectrograms and discrete symbolic scores through a Transformer encoder and a hierarchical decoder.}
  \label{fig:model}
\end{figure*}

\section{Related Work}

Early approaches to A2S transcription predominantly treated the problem as a pipeline of isolated subtasks, such as multi-pitch estimation followed by rhythm quantization~\cite{cogliati2016transcribing,nakamura2018towards,shibata2021non}. While functional, these decomposition-based methods often suffered from error accumulation across different stages. To mitigate this, the field has seen a notable paradigm shift towards end-to-end models~\cite{carvalho2017towards}.

The current standard for end-to-end A2S heavily relies on a hybrid convolutional recurrent neural network (CRNN) architecture within an encoder-decoder framework. Typically, a CNN front-end is deployed to extract local acoustic features from audio spectrograms, which are subsequently decoded into musical symbols by an RNN~\cite{zeng2024end,alfaro2024transformer,arroyo2022neural,martinezsevilla23_interspeech}. Training strategies for these sequence labeling tasks generally diverge into two paths: utilizing Connectionist Temporal Classification (CTC) loss~\cite{roman2018end,arroyo2022neural} to bypass explicit input alignment, or adopting a Sequence-to-Sequence (Seq2Seq) framework~\cite{liu2021joint} for direct target prediction. Despite improving overall transcription coherence, translating dense polyphonic music into structured score formats (e.g., \texttt{ABC notation}, \texttt{**Kern}) remains a formidable challenge, with many models struggling to consistently maintain voicing structures across complex, unconstrained musical textures~\cite{arroyo2022neural}.

To address the hierarchical complexity of musical scores, recent research has explored more advanced decoding mechanisms. Zeng et al.~\cite{zeng2024end} introduced a Seq2Seq model with a hierarchical decoder capable of capturing both bar-level metadata and note-level sequences. To counteract the scarcity of real-world data, they pre-trained their model using an Expressive Performance Rendering (EPR) system to humanize the rendered data before fine-tuning on human piano recordings. Concurrently, recognizing the inherent limitations of RNNs in modeling long-term polyphonic dependencies, Alfaro-Contreras et al.~\cite{alfaro2024transformer} integrated a Transformer decoder into the A2S pipeline. They employed a specialized two-dimensional positional encoding to preserve frequency-time relationships from the CNN feature maps, achieving significant improvements in transcribing polyphonic string music.

Despite these architectural innovations, existing end-to-end models remain fundamentally constrained by the scarcity of large-scale, diverse training data, heavily restricting their applicability to single-instrumentation scenarios. Furthermore, the persistent reliance on CNN acoustic front-ends introduces architectural complexity.

\section{Methodology}\label{sec:methodology} 

In this section, we detail the methodology behind \tutti. Building upon recent advancements in symbolic music processing and generation~\cite{wu2024melodyt5, wang2025notagen}, we adapt the hierarchical bar-stream patching strategy to the challenging cross-modal domain of A2S transcription. We first outline the structured ABC notation adopted for data representation (Section \ref{subsec:data_rep}). We then detail the core architectural design of \tutti, highlighting our continuous acoustic encoder and the cross-modal hierarchical decoder.

\subsection{Data Representation}\label{subsec:data_rep}
We use ABC notation for symbolic music representation. Departing from the original plain ABC format, we adopt an advanced bar-stream patching strategy, along with an interleaved and reduced ABC notation format, to ensure a highly structured and sequence-efficient representation, as established in previous works~\cite{wu2023tunesformer, wu2023clamp, wang2025notagen}.

\subsection{Model Architecture}
As depicted in Figure \ref{fig:model}, \tutti employs an encoder-decoder architecture based on the Transformer network~\cite{vaswani2017attention}, tailored specifically for A2S transcription. 

The model directly ingests audio spectrograms and autoregressively generates hierarchical symbolic patches. It is composed of the following key modules:

\textbf{Acoustic Adapter:} The encoder's input consists of continuous acoustic features. Specifically, raw audio is first transformed into a spectrogram representation with the shape of $T \times F$, where $T$ represents the number of time steps and $F$ denotes the frequency bins. To align these high-dimensional acoustic features with the internal working dimension of the Transformer, we apply a dedicated linear projection layer. This layer maps the 1025-dimensional vector at each time step into a dense continuous embedding of dimension 512, matching the model's hidden size ($d_\text{model}$). This projection acts as a lightweight acoustic embedding mechanism, compressing the frequency information while preserving the temporal sequence length, enabling the Transformer to directly ``read'' the audio.

\textbf{Acoustic-level Encoder:} Operating on the sequence of continuous embeddings produced by the projection layer, the encoder is responsible for extracting high-level acoustic and harmonic contexts. The encoder captures both local transient events (e.g., note onsets) and global structural dependencies (e.g., phrasing and sustained pitches) directly from the spectrogram sequence over the $T$ time steps.

\textbf{Cross-Modal Patch-level Decoder:} To handle the highly structured and dense nature of musical scores, we adopt the hierarchical decoding strategy~\cite{wu2024melodyt5}, tailoring it for cross-modal alignment. This module takes the embeddings of previously encoded acoustic features as input. This allows the patch-level decoder to dynamically align the global musical structure (e.g., bar-level pacing and polyphony) with corresponding acoustic segments in the audio sequence, ultimately outputting a dense contextual representation for the next predicted bar.

\textbf{Character-level Decoder:} Operating in a cascaded, step-by-step manner, this module acts as the final translator into discrete ABC notation. Taking the dense contextual representation generated by the patch-level decoder as its condition vector, it autoregressively decodes the specific musical tokens (characters) within that patch. By focusing on local syntactic information—such as specific pitches, durations, and structural symbols—it sequentially reconstructs the exact musical content of each bar patch until the full target score is generated. This hierarchical design effectively decouples the global audio-to-measure alignment from the local measure-to-character generation, ensuring both structural coherence and syntactic correctness.

\section{Dataset}
To overcome the severe scarcity of high-quality, real-world data, we introduce a massive, fully synthesized multi-instrumentation \tutticorpus dataset. A fundamental contribution of this work is the creation of this dataset from scratch: rather than relying on existing, heavily constrained symbolic repositories, we leverage NotaGen~\cite{wang2025notagen}---a symbolic music generation model---to generate 363,610 unique ABC scores, conditioned on diverse prompts specifying musical periods (e.g., Baroque, Romantic), composers (e.g., Bach, Beethoven), and instrumentations (e.g., Keyboard, Chamber). This ensures stylistic variety and musical coherence, as NotaGen's fine-tuning on high-quality classical corpora and RL alignment enhance generation quality. To mitigate potential data leakage from NotaGen's training distribution, we conducted a rigorous similarity audit between our generated corpus and the ground-truth test set. We computed the normalized Levenshtein (edit) distance over canonicalized musical sequences to ensure that no generated piece duplicates or closely resembles any test sample. Any synthetic instance exceeding a minimal similarity threshold was strictly removed, guaranteeing that the evaluation benchmark remains completely unseen.

\begin{figure*}[t]
  \centering
  
  \begin{minipage}[c]{0.48\textwidth}
    
    \centering
    \small
    \begin{tabular}{llrr}
      \toprule
      \textbf{Category} & \textbf{Details} & \textbf{Count} & \textbf{\%} \\
      \midrule
      \textbf{Total Segments} & All audio-score pairs & \textbf{363,610} & 100.00 \\
      \midrule
      \textbf{Top 5} & Piano & 228,964 & -- \\
      \textbf{Instruments} & Violin & 208,119 & -- \\
      \textit{(by frequency)} & Cello & 102,367 & -- \\
      & Harpsichord & 63,974 & -- \\
      & Viola & 44,912 & -- \\
      \midrule
      \textbf{Ensemble} & Solo (1 Inst.) & 235,032 & 64.64 \\
      \textbf{Formats} & Duet (2 Inst.) & 51,049 & 14.04 \\
      \textit{(by size)} & Trio (3 Inst.) & 53,601 & 14.74 \\
      & Quartet (4 Inst.) & 14,010 & 3.85 \\
      & Quintet (5 Inst.) & 5,522 & 1.52 \\
      & Large Ensemble & 4,396 & 1.21 \\
      \bottomrule
    \end{tabular}
    \captionof{table}{Detailed statistics of the \tutticorpus dataset, highlighting instrumentation frequency and ensemble formats.}
    \label{tab:dataset_stats}
    \vspace{6pt} 
  \end{minipage}%
  \hfill
  \begin{minipage}[c]{0.48\textwidth}
    \centering
    
    \includegraphics[width=\textwidth]{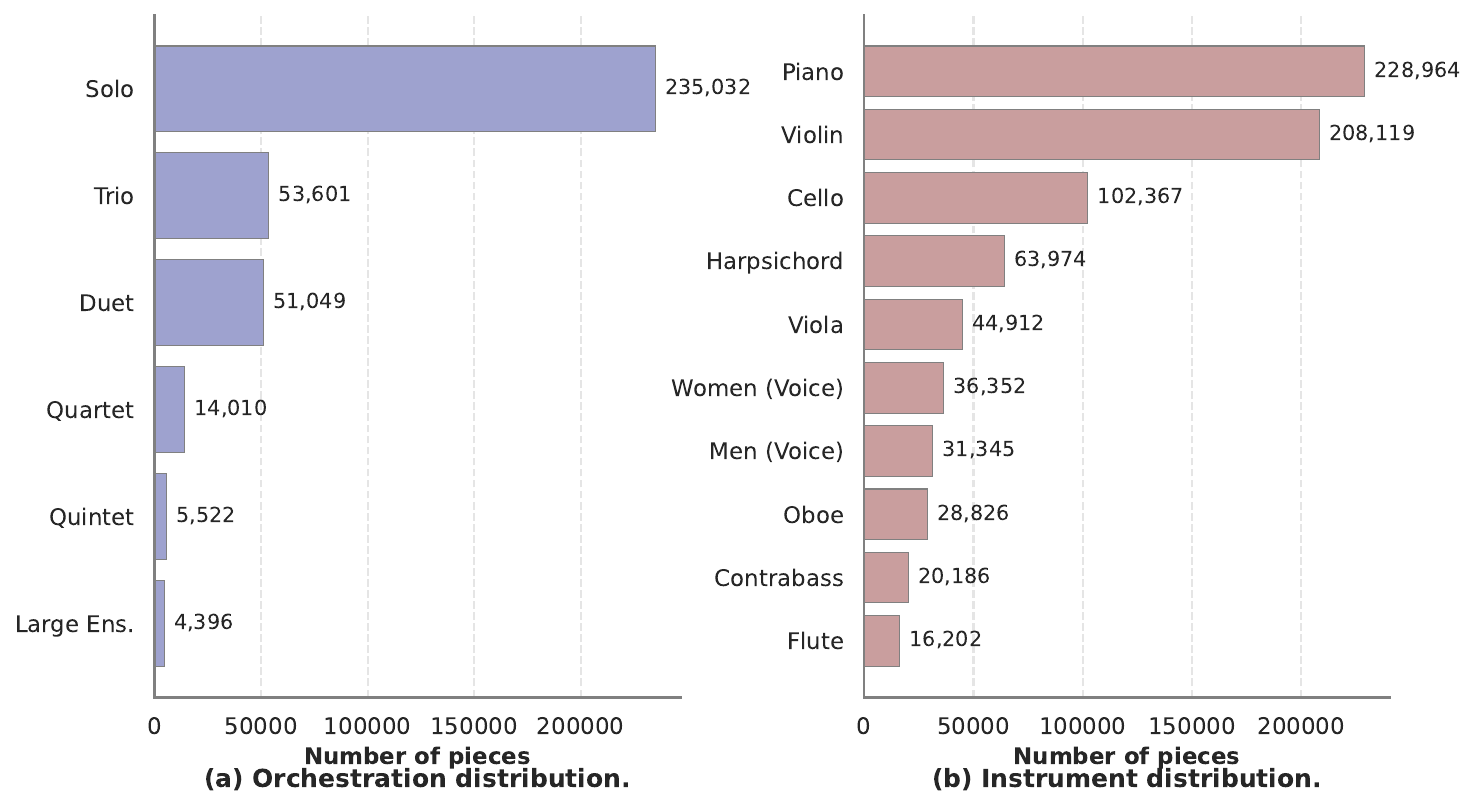}

    \vspace{6pt} 
    \caption{Distributions of ensemble types (left) and individual instruments (right) across the \tutticorpus dataset.}
    \label{fig:dataset_dist}
  \end{minipage}
  
\end{figure*}

\subsection{Data Synthesis and Alignment Pipeline}

\subsubsection{Score Standardization and Compaction}
Raw ABC notation files often contain heterogeneous metadata, varying repeat structures, and empty instrumental voices, which introduce noise during sequence modeling. We first standardize the ABC files by removing redundant metadata, stylistic textual annotations, and unplayable symbols. Crucially, we unroll all repeat structures to ensure the structural linearity of the score matches the temporal linearity of the audio sequence. Furthermore, we dynamically detect and strip empty voices, transferring necessary instrumental metadata to active voices, and compress the multi-line voice structures into a dense, inline sequence representation. This compact format maximizes the efficiency of the Transformer model during training.

\subsubsection{Expressive Performance Rendering (EPR)}

A major challenge in A2S transcription is the discrepancy between rigid, quantized synthetic audio and human expressive variations. To improve model generalization~\cite{zeng2024end}, we apply Expressive Performance Rendering (EPR) prior to audio synthesis via three parallel pipelines. A Plain pipeline applies strict quantization with baseline dynamics compression and voice balancing. For piano tracks, we utilize VirtuosoNet~\cite{jeong2019virtuosonet} to generate human-like expressive performances natively. Finally, since neural EPR is largely limited to pianos, we introduce a Rule-based EPR for other instruments that injects controlled perturbations, including velocity humanization ($\pm 15$), timing jitter ($\pm 10$ ms), and tempo rubato ($\pm 3\%$).

\subsubsection{Audio Synthesis and High-Precision Alignment}

Instead of relying on standard General MIDI (GM) synthesizers, which lack acoustic nuances, we render the processed MIDI files into high-quality audio using the SFZ sample format via sfizz.\footnote{https://github.com/sfztools/sfizz} We extract the required pitch ranges from the ABC score to dynamically match each voice with the most appropriate acoustic sample library. The resulting individual audio tracks are mixed with linear overlay and normalized to a target loudness of -23.0 LUFS.

To establish ground-truth alignment for segmentation, we employ Dual Dynamic Time Warping (DualDTW)~\cite{peter-offline2023} via Parangonar\footnote{https://github.com/sildater/parangonar} to align the expressive MIDI performances back to the quantized MusicXML scores. This yields precise, bar-level temporal anchors. Using these anchors, we greedily segment both the continuous audio and the inline ABC scores into aligned chunks no longer than 14.8 seconds.

Finally, the segmented audio clips are converted into power spectrograms (sample rate 22,050 Hz, window size 2048, hop length 160) and log-dB normalized, serving as the direct input to our model.

\subsection{Data Statistics}
Through our automated generation and synthesis pipeline, we successfully compiled a massive, multi-instrumentation dataset comprising 363,610 unique and fully aligned data pairs. Our dataset introduces a high level of instrumentation and compositional diversity, making it highly suitable for training generalized transcription models.

As detailed in Table \ref{tab:dataset_stats} and Figure \ref{fig:dataset_dist}, the dataset covers an extensive range of instruments across different families, including strings, keyboards, and various woodwinds/brass. In terms of instrumentation configurations, the dataset naturally follows a long-tailed distribution characteristic of diverse musical compositions. While solo configurations account for 64.6\% of the data---providing a solid foundation for fundamental pitch and duration recognition---a substantial portion of the dataset is dedicated to complex polyphonic and multi-instrumentation combinations. Specifically, duets and trios constitute nearly 29\% of the corpus, and approximately 24,000 instances feature quartets, quintets, or larger ensembles. This structural diversity forces the model to learn generalized acoustic representations and robust source identification capabilities, rather than overfitting to the timbre of a single instrument.

\begin{table*}[t]
\centering
\small
\begin{tabular}{cccccccc}
\toprule
\multirow{2}{*}{\textbf{Dataset}} & \multirow{2}{*}{\textbf{Method}} & \multicolumn{6}{c}{\textbf{Metrics (MV2H)}} \\
\cmidrule(lr){3-8}
 & & Multi-pitch & Voice & Meter & Harmony & Note Value & Total \\
\midrule
\multirow{4}{*}{ASAP} & Full Instrumentation Pre-training & \textbf{70.7} & 87.5 & \textbf{90.9} & 55.6 & \textbf{90.7} & \textbf{76.1}$^{\dagger}$ \\
 & Single Instrumentation Pre-training & 66.2 & 86.7 & 90.6 & \textbf{57.1} & 90.1 & 75.0$^{\dagger}$ \\
 & Without Pre-training & 13.2 & 74.2 & 78.4 & 54.6 & 71.5 & 53.4$^{\dagger}$ \\
 & \textbf{Zeng et al.}~\cite{zeng2024end} & 63.3 & \textbf{88.4} & -- & 54.5 & \textbf{90.7} & 74.2$^{\dagger}$ \\
\midrule
\multirow{4}{*}{Quartets} & Full Instrumentation Pre-training & \textbf{97.7} & \textbf{99.5} & \textbf{97.8} & 83.4 & \textbf{99.7} & \textbf{95.6} \\
 & Single Instrumentation Pre-training & 93.8 & 98.4 & 95.3 & 83.4 & 99.7 & 93.0 \\
 & Without Pre-training & 46.2 & 85.5 & 82.9 & 56.0 & 92.0 & 72.5 \\
 & \textbf{Alfaro-Contreras et al.}~\cite{alfaro2024transformer} & 70.6 & 92.1 & 70.7 & \textbf{97.6} & 93.4 & 84.9 \\
\midrule
\multirow{3}{*}{Tenor Saxophone} & Full Instrumentation Pre-training & \textbf{80.4} & \textbf{100.0} & \textbf{96.1} & 51.8 & \textbf{98.0} & \textbf{85.3} \\
 & Without Pre-training & 26.9 & 92.3 & 85.9 & 26.9 & 46.8 & 55.8 \\
 & {\textbf{Martínez-Sevilla et al.}}~\cite{martinezsevilla23_interspeech} & 48.1 & 65.2 & 61.3 & \textbf{63.0} & 64.5 & 60.3 \\
\midrule
\multirow{3}{*}{Alto Saxophone} & Full Instrumentation Pre-training & \textbf{82.8} & \textbf{100.0} & \textbf{96.6} & 46.5 & \textbf{98.7} & \textbf{84.9} \\
 & Without Pre-training & 23.4 & 91.7 & 79.3 & 20.8 & 47.4 & 52.5 \\
 & {\textbf{Martínez-Sevilla et al.}}~\cite{martinezsevilla23_interspeech} & 40.1 & 67.3 & 57.6 & \textbf{67.2} & 67.1 & 59.9 \\
\bottomrule
\end{tabular}
\caption{Performance comparison (MV2H scores~\cite{McLeod:18a}) between \tutti and baseline models across multiple datasets. \textbf{Bold} values indicate the best performance in each category.$^{\dagger}$Total score is calculated as the average of four available metrics.}
\label{tab:results}
\end{table*}

\section{Experiment}

\subsection{Settings}
The experiments are designed to systematically assess the capabilities and generalizability of \tutti in various A2S tasks. We utilize the \tutticorpus dataset, which is randomly split into 95\% for training and 5\% for testing.

\tutti features a 9-layer patch-level encoder and decoder, a 3-layer character-level decoder, and a hidden size of 512, amounting to 88.9M parameters. A linear projection layer maps the 1025-dimensional spectrogram features into the encoder's input space. The model processes audio inputs of up to approximately 14.8 seconds using a patch length of 2048. It employs a 128-size ASCII-based vocabulary, using characters 0--2 for special tokens (pad, bos, and eos).

The AdamW optimizer~\cite{loshchilov2017decoupled} is used with a learning rate of 2e-4 for pre-training and 1e-5 for fine-tuning. Pre-training includes a 1000-step warmup followed by a constant learning rate over 32 epochs, with a batch size of 10 per GPU. Pre-training was conducted on 8 H800 GPUs, taking approximately 6 days to complete. Fine-tuning was performed on 6 H800 GPUs with a batch size of 6 per GPU over 32 epochs.

In ablation studies, we investigate the effects of multi-instrumentation learning on \tutti, considering three settings: 1)
omitting pre-training, 2) using only the downstream task-specific data from \tutticorpus, or 3) utilizing the entire
\tutticorpus, which includes all instrumentations.

In terms of comparative evaluations, we select open-source models that excel in their respective domains for benchmarking. \tutti is fine-tuned on identical datasets to these models, ensuring fairness in comparison.

\subsection{Ablation Studies}
To better understand the contributing factors to our model's performance, we conduct ablation studies focusing on two main aspects: 1) the necessity of our large-scale synthetic pre-training paradigm, and 2) the effectiveness of multi-instrumentation representation learning compared to instrumentation-specific pre-training.

As depicted in Table \ref{tab:results}, the model trained strictly on the target datasets experiences a severe performance degradation, falling behind the baseline models across all tasks. This aligns with the well-known data-hungry nature of Transformer architectures~\cite{kaplan2020scaling,shao2022transformers}. Our results clearly demonstrate that without the proposed large-scale synthetic data pre-training, the Transformer struggles to generalize on limited real-world target data. Thus, our synthetic pre-training is the critical enabler that unlocks the high capacity of the Transformer for A2S tasks. While our experiments demonstrate that large-scale pre-training unlocks the potential of purely Transformer-based models, we acknowledge that this data-centric paradigm is architecture-agnostic; CNN-based hybrid models would likely exhibit parallel performance gains if exposed to the same scale of pre-training.

We further investigate whether pre-training on a diverse mixture of instrumentations provides an advantage over pre-training on a matched, single-instrumentation distribution. In the ``single instrumentation pre-training'' setting, we restricted the synthetic pre-training data to match the target task (e.g., using only synthetic piano data for the ASAP piano task, and synthetic string quartets for the Quartets dataset). Note that for the saxophone datasets, this ablation is not applicable since saxophones were entirely excluded from our pre-training corpus.

The results demonstrate that the full-instrumentation pre-trained model outperforms the single-instrumentation pre-trained counterpart. This indicates that multi-instrumentation pre-training is not merely about covering more instrument types, but it fundamentally enhances the model's underlying representation capability. By learning to disentangle complex, multi-timbral polyphony across diverse instruments, the model learns more robust universal acoustic features and harmonic relationships, which ultimately benefits its performance even on instrument-specific downstream tasks.

\subsection{Comparative Evaluations}
We compare \tutti, a generalizable A2S model pre-trained on the \tutticorpus dataset, with several baseline models. The following baseline models have been selected for comparison:
\begin{itemize}
    \item {\textbf{Zeng et al.}}~\cite{zeng2024end} employ a convolutional recurrent neural network (CRNN) with a customized hierarchical decoder and the \textit{ASAP} (piano)~\cite{asap-dataset} dataset as fine-tuning data.
    \item {\textbf{Alfaro-Contreras et al.}}~\cite{alfaro2024transformer} utilize a CNN with 2-D positional encoding and a Transformer decoder trained on the string quartet dataset.
    \item {\textbf{Martínez-Sevilla et al.}}~\cite{martinezsevilla23_interspeech} use a CRNN trained on human-recorded saxophone data. As the saxophone is absent from the \tutticorpus instrumentation, this comparison serves to evaluate the model's transfer learning capability on unseen instruments.
\end{itemize}

To ensure a comprehensive assessment, we adopt the MV2H metric~\cite{McLeod:18a}, which comprises five sub-metrics: multi-pitch detection, voice separation, metrical alignment, harmonic detection, and note value accuracy. These values are averaged to produce the total MV2H score. Note that for Zeng et al.~\cite{zeng2024end}, the metrical alignment metric is unavailable; thus, its total score is calculated by averaging the remaining four components.

Table \ref{tab:results} presents the quantitative results. Overall, \tutti pre-trained with full instrumentation outperforms all baselines, demonstrating superior transcription capabilities across both polyphonic and monophonic tasks. On the ASAP dataset, which involves expressive timing and complex polyphony, our method achieves a significant improvement in multi-pitch accuracy, outperforming the baseline by 7.4 points. While yielding competitive results in voice tracking, \tutti achieves a higher total score, proving its robustness in handling expressive real-world piano recordings.

For the Quartets dataset, our model demonstrates substantial margins over the baseline across almost all metrics, increasing the total score from 84.9 to 95.6. The near-perfect scores in voice (99.5) and note value (99.7) indicate that our approach reliably transcribes dense multi-track arrangements.

A key strength of \tutti is its transfer learning capability. On the Saxophone datasets, our model successfully generalizes to these unseen instruments despite their total absence from the \tutticorpus training set. Notably, since the saxophone is a monophonic instrument, voice accuracy serves as a diagnostic indicator of whether the model correctly identifies its single-voice nature. Failing to reach 100\% implies the model erroneously predicts multiple simultaneous voices---a hallucination likely caused by insufficient acoustic understanding of the instrument. Our fully pre-trained model achieves this theoretical ceiling (100.0) on both datasets, whereas the task-specific baseline and the non-pre-trained variant do not. We attribute this to the diverse multi-timbral representations learned during pre-training: by exposure to a wide variety of instruments and ensembles, \tutti acquires a more robust acoustic grounding that generalizes to unseen timbres without overfitting to a single instrument's characteristics.

\section{Conclusion}
We introduced \tutti, a unified, Transformer encoder-decoder framework that addresses data scarcity and architectural complexity bottlenecks in generalizable A2S transcription. By pre-training on \tutticorpus---our large-scale dataset comprising over 363,000 synthetic multi-instrumentation audio-score pairs---we demonstrated that a data-centric paradigm can effectively compensate for the absence of local inductive biases typically provided by CNN acoustic front-ends.

Extensive objective evaluations reveal that \tutti outperforms established task-specific baselines across diverse polyphonic and monophonic tasks, establishing new overall state-of-the-art results. Crucially, the model exhibits remarkable transfer learning capability, achieving theoretical voice accuracy on unseen real-world instruments, such as the saxophone, strictly through the robust representational power learned from synthetic data.

Finally, we release the \tutticorpus dataset as a rich and highly scalable resource to propel future research in unified A2S transcription. Although \tutti marks a significant step forward, extending our mechanism to seamlessly handle unconstrained, full-piece transcriptions remains a promising avenue for future work.

\clearpage
\section{Acknowledgments}
This work was supported by the following funding source: Special Program of National Natural Science Foundation of China (Grant No. T2341003).

\bibliography{ISMIRtemplate}

\end{document}